\documentclass[aps,twocolumn,prl,superscript,floatfix,superscriptaddress,showpacs,footinbib]{revtex4-1}

\usepackage{amssymb}

\usepackage[pdftex]{graphicx}
\usepackage{dcolumn}
\usepackage{bm}
\usepackage{amsmath}
\usepackage{array}
\usepackage{color}
\usepackage{float}
\usepackage{subfigure}
\usepackage{dsfont}
\usepackage{txfonts}
\usepackage{wasysym}
\usepackage{multirow}
\usepackage{sidecap}
\usepackage{xcolor,cancel}
\usepackage[normalem]{ulem}
\usepackage{natmove}  

\usepackage{hyperref}
\hypersetup{
    colorlinks,%
    citecolor=blue,%
    linkcolor=blue,%
    urlcolor=blue
}

\begin{document}

\title{Competing long-range  interactions and spin vector chirality in spin-chains}

\author{E. Vernek}
\affiliation{Instituto de F\'isica, Universidade Federal de 
Uberl\^andia, Uberl\^andia, Minas Gerais 38400-902, Brazil}
 
\affiliation{Department of Physics and Astronomy, and Nanoscale and 
 Quantum
Phenomena Institute, Ohio University, Athens, Ohio 45701-2979, USA}

\author{O. \'Avalos-Ovando}

\affiliation{Department of Physics and Astronomy, and Nanoscale and 
	Quantum
	Phenomena Institute, Ohio University, Athens, Ohio 45701-2979, USA}

\author{S. E. Ulloa}

\affiliation{Department of Physics and Astronomy, and Nanoscale and 
	Quantum
	Phenomena Institute, Ohio University, Athens, Ohio 45701-2979, USA}

\date{\today}

\begin{abstract}
Studies of competing orders in 1D magnetic chains have attracted considerable attention in
recent years, as the presence of long-range Heisenberg interactions is found to allow interesting quantum phase transitions. We investigate here the role of spin-orbit effects by considering spin-1/2 chains in the presence of both collinear and non-collinear long-range interactions. By employing  exact diagonalization and density matrix renormalization group calculations we investigate the  rich phase diagram of this system. We find transitions from collinear to transverse magnetic correlated order as the strength of the non-collinear coupling increases, accompanied by a jump in the vector spin chirality order parameter of the system. %
This shows that tuning long-range interactions allows control of the onset of sizable vector spin chirality in a system, which may be used to transmit information down the chain.
We investigate the characteristic structure of the distinct phases and explore their possible physical implementation in different materials systems.
\end{abstract}
\maketitle

Characterization of long-range correlations in low dimensional systems has attracted a great deal of attention since the birth of quantum mechanics, resulting in a substantial body of work~\cite{Book_Auerbach,Book_Giamarchi}. More recently, correlations have gained renewed interest as they are considered a key ingredient in quantum technology requiring long distance entanglement \cite{Bazhanov2018}. Long-range interactions are essential to overcome the celebrated Mermin-Wagner theorem~\cite{PhysRevLett.17.1133} that rules out long-range order (LRO) at finite temperature in 1D and 2D systems with short-range interactions~\cite{Dimensions1967,Bruno2001}. A natural platform to explore long-range quantum correlations is provided by magnetic chains in different systems, many of which exhibit long-range interactions~\cite{PhysRevLett.91.207901,PhysRevLett.99.060401, Sahling2015,PhysRevLett.96.247206,Wei2018FLUORAPATITE}. Indeed, much theoretical effort has been devoted to understanding the quantum phases of strongly correlated spin chains~\cite{Cardy_1981,Affleck_1989}. 

An interesting class of correlated systems include antiferromagnetically coupled spin chains with long-range interactions \cite{doi:10.1063/1.1664978,doi:10.1063/1.1664979,OKAMOTO1992433} in which the spin configuration of the ground state does not in general minimize the energy locally, rendering a rich phase diagram in such frustrated system~\cite{Diep}. Sandvik~\cite{Sandvik2010} has explored the phase diagram of a related frustrated spin chain with long-range collinear interactions that decay as $1/r^\alpha$, where $r$ is the distance between spins and  $\alpha> 0$. He finds a first-order transition between the dimerized phase for short range interaction and a long-range ordered state as the frustration interaction increases.

Another interesting example of a complex system occurs when a slow decaying Ruderman-Kittel-Kasuya-Yosida (RKKY) effective exchange competes with the Dzyaloshinskii-Moriya (DM) interaction in a magnetic system.  The RKKY collinear interaction (proportional to ${\bf S}_i\cdot {\bf S}_j$, where ${\bf S}_l$ represent different spins operators in the system) tends to align the spins anti/parallel to each other.  In contrast, the non-collinear DM coupling  is proportional to ${\bf S}_i\times {\bf S}_j$, favoring a relative perpendicular alignment. As a result, the system may possess a frustrated ground state with rich three-dimensional spin structures that include skyrmions \cite{Fert2017REVIEW,EverschorSitte2018REVIEW}.
DM interactions between magnetic impurities can be seen to derive from spin-orbit coupling (SOC) in the carriers of the host \cite{DZYALOSHINSKY1958241,PhysRev.120.91,Nagaosa2010RevModPhys}. 

While the RKKY interaction is well known to behave as $\cos{(2 k_F r)}/r^d$, with the Fermi momentum $k_{F}$ and $d$ the dimensionality of the host electron system, the combination of confinement, orbital content, and/or the effect of SOC  results in small $d$ values and consequently long-range interactions.  This is especially seen in 2D materials, 
including graphene \cite{Black2010,Duffy2014,Zhang2017VeselagoLens}, silicine~\cite{Zare2016}, and transition metal dichalcogenides~\cite{PhysRevB.93.161404,JPCMAvalosOvando2018,PhysRevB.99.035107}, as well as in 1D quantum wires with SOC, which show effective $d \lesssim 1$ behavior~\cite{Silva_2019}.
In graphene nanoribbons, for instance, while impurities in the middle of the ribbon interact as $r^{-2}$,  those along zigzag~\cite{Black2010} and armchair~\cite{Duffy2014} edges exhibit a power-law decay with $1<d<2$ for small  distances. Magnetic impurities in graphene $p$-$n$ junctions also show enhanced range and strength of RKKY interactions \cite{Zhang2017VeselagoLens}, similar to interfaces in topological insulators surfaces~\cite{PhysRevB.99.195456}, and in transition metal dichalcogenides, where the effective interaction strength and range can be tuned by gating, such that the interaction decays with $d \lesssim 1$ \cite{PhysRevB.93.161404,JPCMAvalosOvando2018,PhysRevB.99.035107}. 

Recent elegant experiments have also probed the long-range exchange interaction of magnetic impurities adsorbed on different metal surfaces, and the importance of DM interactions in determining the appearance of a spin-spiral ground state configuration \cite{Steinbrecher2018,Hikihara2008}.  Those experiments highlight the appearance of a vector spin chirality (VSC) $ {\bf \kappa}_j \simeq \langle {\bf S}_{j-1} \times {\bf S}_j \rangle$, averaged over the state of the spin chain, and identified as an order parameter.  VSC points to the importance of DM interaction in the system dynamics, and interestingly, changing it under external driving field may be used to transmit information down the chain \cite{Katsura2005,Menzel2012}.  Understanding to what extent  VSC is established in a system is an interesting question, especially as strong DM interactions may be present in real systems.

\begin{figure}[!t]
	\subfigure{\includegraphics[width=3.5in]{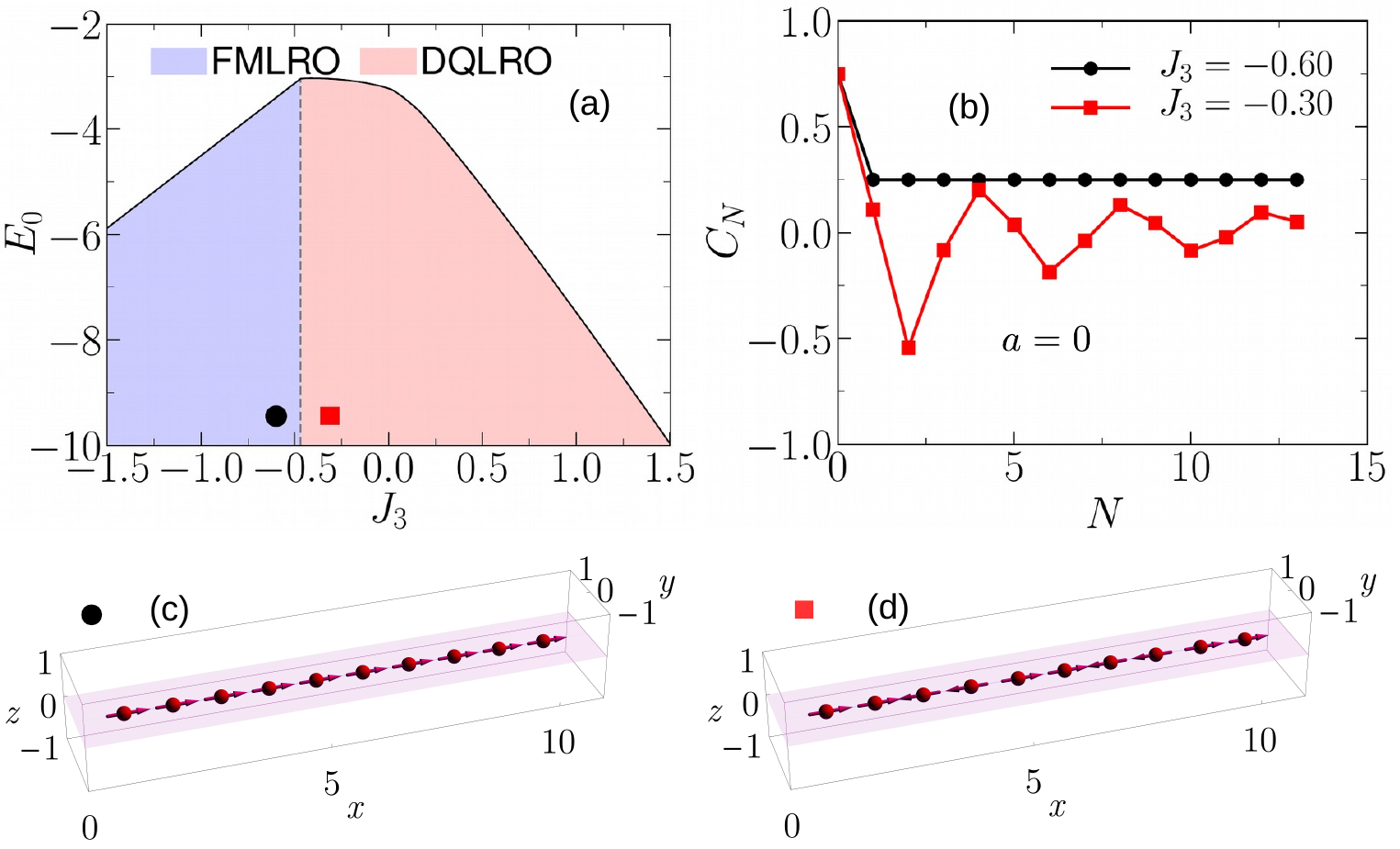}}   
	\caption{(a) Ground state energy as a function of $J_3$ with no DM interactions ($a=0$, $L=14$). Blue (and red) shading indicates a ferromagnetic (and dimerized) phase in the system. (b) Longitudinal correlation between the first and the $N$-th spin of the chain. Black and red curves correspond to $J_3$ on different phases marked by a black circle and a red square in panel (a). (c) and (d) Panoramic view  of the magnetic moments in the ferromagnetic and dimerized phases, respectively.} 
	\label{fig1}
\end{figure}

Motivated by these possibilities, we consider here a system composed  of an open chain of spin-$1/2$ impurities with both RKKY and DM  interactions.
While several studies  on 1D systems with long-range RKKY interactions have been reported, the effect of long-range  DM interactions has been explored  much less. In particular, it is natural to ask how  DM interactions modify the different phases of the collinear model and how they may affect the existence of frustrated ground states of the system.  
We find that as DM couplings are turned on, there is a gradual evolution into either a `tilted ferromagnetic' or `tilted dimerized' phase, with spins rotating slightly along the chain; both phases having negligible VSC.\@  However, as DM interactions grow stronger, there is a transition  into a phase with cycloidal quasi-long range order, where the VSC is seen to jump significantly, signaling the onset of a drastically different phase with distinct spin correlations.  

The Hamiltonian describing the system is given as
\begin{eqnarray}\label{H0}
H=\sum_{i\neq j \in \{ L \} }
J_{ij} \, {\bf S}_i\cdot {\bf S}_j +D_{ij} 
\left({\bf S}_i\times {\bf S}_j \right)_z \, , 
\end{eqnarray}
where $J_{ij}$ and $D_{ij}$ represent the RKKY and DM interaction strengths and $L$ is the number of spins in the chain. Notice that $J_{ij}> 0$ favors antiparallel (antiferromagnetic) alignment (and ferromagnetic for $J_{ij}<0$). For simplicity we  consider uniform couplings up to third nearest neighbor, $J_{i, i+1}\equiv J_1$, $J_{i,i+2}\equiv J_2$, and $J_{i,i+3}\equiv  J_3$.

Notice that for  $D_{ij}=0$ and $J_3=0$, this model can describe  the so called $J_1$-$J_2$ frustrated model which has a dimerized ground state for $J_2 \gtrsim 0.241 J_1 >0$ \cite{OKAMOTO1992433,PhysRevB.54.R9612}. We consider a different regime by fixing $J_1 \equiv J>0$ and $J_2\equiv -J/2$. In this case, if $J_3 < J_{3c}\approx -0.49 J$, the ground state corresponds to a ferromagnetic (FM) ordered phase, while for $J_3 > J_{3c}$ the ground state exhibits a dimerized quasi-long range order (DQLRO), similar to the phase with spatial period $\pi/2$ reported for large and positive $J_2$ \cite{Sandvik2010}. 

To investigate the ground state properties of Hamiltonian \eqref{H0} we combine exact diagonalization using the QuSpin package \cite{Weinberg2017} and density matrix renormalization group (DMRG) calculations  using the ITensor library \cite{ITensor}. While the first is  convenient for small chains, the latter allows us to investigate much larger systems.  We use $J=1$ as the energy scale, so that $J_1=1$, $J_2=-\frac{1}{2}$, and consider different values of $J_3$. The DM interactions are set as $D_{i,j+\ell}\equiv D_{\ell}=a J_{\ell}$ for $\ell=1,2,3$, so that $a$ represents the overall scaling of DM effects.

In Fig.\ \ref{fig1} we set $a=0$, so that DM couplings do not take place; \ref{fig1}(a) shows the ground state energy  of the system ($E_0$) as a function of $J_3$. A sudden change in the behavior of $E_0$ is seen at $J_3=J_{3c} \approx -0.49 $. Easily identified as a level crossing,  this marks the transition between two distinct phases in the system; a FM phase with LRO (blue) for $J_3 < J_{3c}$ and  the dimerized  phase (DQLRO, red, $J_3>J_{3c}$).  Figure \ref{fig1}(b)  shows the ground state longitudinal spin correlations $C_N=\langle {\bf S}_0 \cdot {\bf S}_N  \rangle$ for two points close to the boundary between the two phases \cite{pinning_field}. The black curve corresponds to  $J_3=-0.6$, as indicated with a black circle in Fig.\ \ref{fig1}(a). Note that $C_N=1/4$ for all sites along the chain, as expected for the ground state with FMLRO.\@ To visualize the magnetic order, we calculate the local magnetization ${\bf M}_N=\langle {\bf S}_N\rangle$  along the chain \cite{pinning_field}. The lower panels give views of the resulting spin structure for different points in the phase diagram. In Fig.\ \ref{fig1}(c) all spins point in the same direction, as the ground state has full FM polarization. For the DQLRO phase, the red line in Fig.\ \ref{fig1}(b) shows $C_N$ for $J_3=-0.3$ [red square in Fig.\ \ref{fig1}(a)].  We observe $C_N$ exhibits rather different behavior, with  oscillations of wavelength $k=2\pi/\lambda \eqsim\pi/2$. This phase is similar to the phase identified as VBS+QLRO($\pi/2$) for large enough $J_2$ in the infinite range model \cite{Sandvik2010}. 
A corresponding view of the local magnetization in this phase is shown in Fig.\ \ref{fig1}(d). The dimerized structure is evident. 
\begin{figure}[t!]
	\centering
	\subfigure{\includegraphics[clip,width=3.35in,height=1.25in]{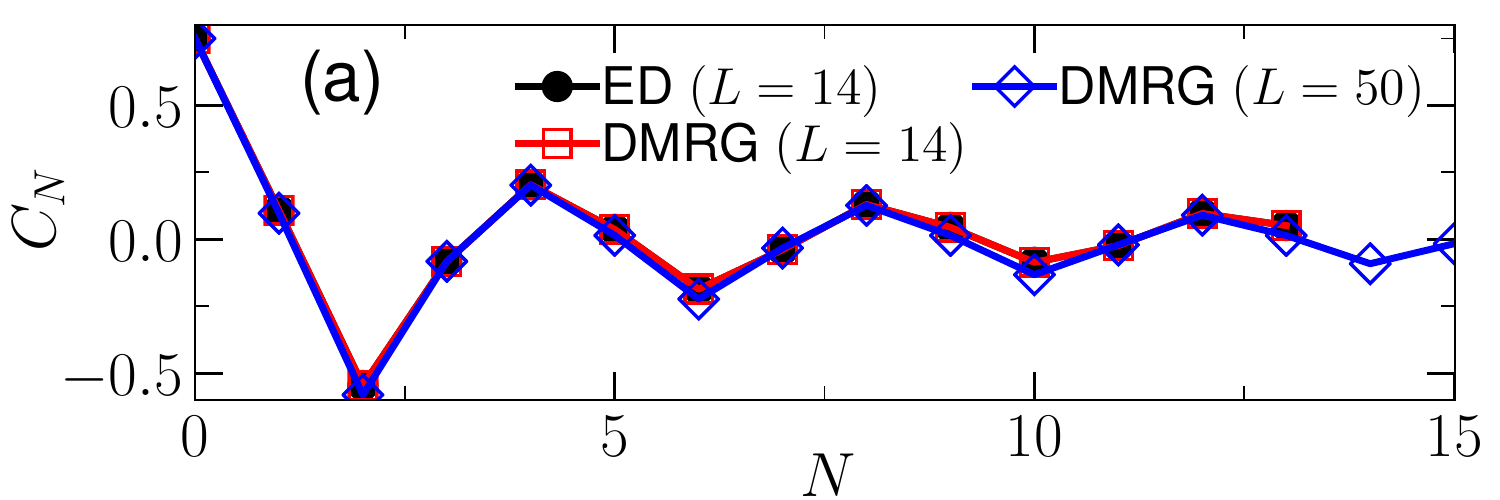}}    
\vskip-0.2cm
	\subfigure{\includegraphics[clip,width=3.35in,height=1.25in]{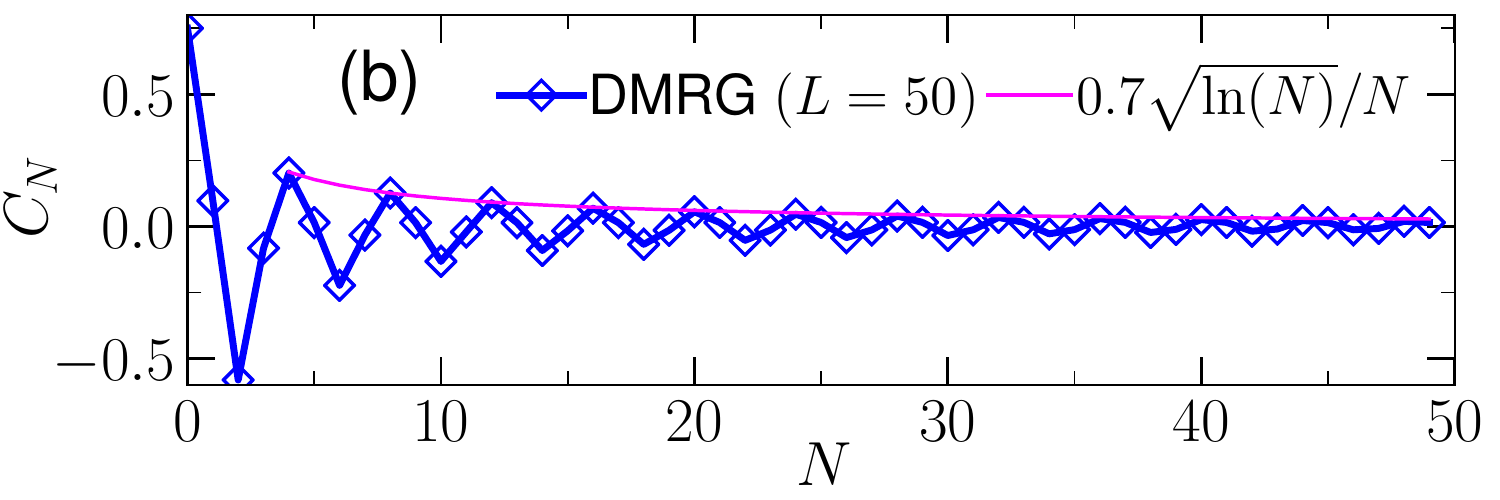}}     
	\caption{(a) Comparison of spin-spin correlations for ED (exact diagonalization) and DMRG calculations  for different chain length $L$. There is remarkable agreement for ED and DMRG for $L=14$. Slight deviation between $L=14$ and $L=50$ results indicate the effect of finite size.  (b) $C_N$ vs $N$ from DMRG calculations for larger distances. Solid magenta line corresponds to $f(N)=0.7\sqrt{\ln(N)}/N$. Here, $a=0$ and $J_3=-0.3$ in both panels, as in Fig.\ \ref{fig1}(b).
	\label{fig4}} 
\end{figure} 

In order to explore system correlations for larger chains we employ a DMRG approach. A comparison of  DMRG and exact diagonalization (ED) results is shown in Fig.\ \ref{fig4}(a), where $C_N$ for $L=14$, using  ED (black dots) and DMRG (red squares) are nearly identical. Blue diamonds correspond to  $L=50$ using DMRG (shown up to $N=15$) and are only slightly different near the minima, likely due to finite size effects. Figure \ref{fig4}(b) exhibits the same results for $L=50$; we note that the decaying oscillating pattern extends far along the entire chain. The solid (magenta) line in \ref{fig4}(b) corresponds to the function $f(x) \simeq \sqrt{\ln{x}}/x$, the behavior expected for the ground state of a chain with only antiferromagnetic nearest neighbor coupling ($J_1>0$)~\cite{Book_Giamarchi}.  That it fits the decay of $C_N$ quite nicely here as well suggests that the introduction of $J_2<0$ and small $J_3$ does not appreciably affect the correlations \cite{Sandvik2010}. 

\begin{figure}[!t]
	\subfigure{\includegraphics[width=3.5in]{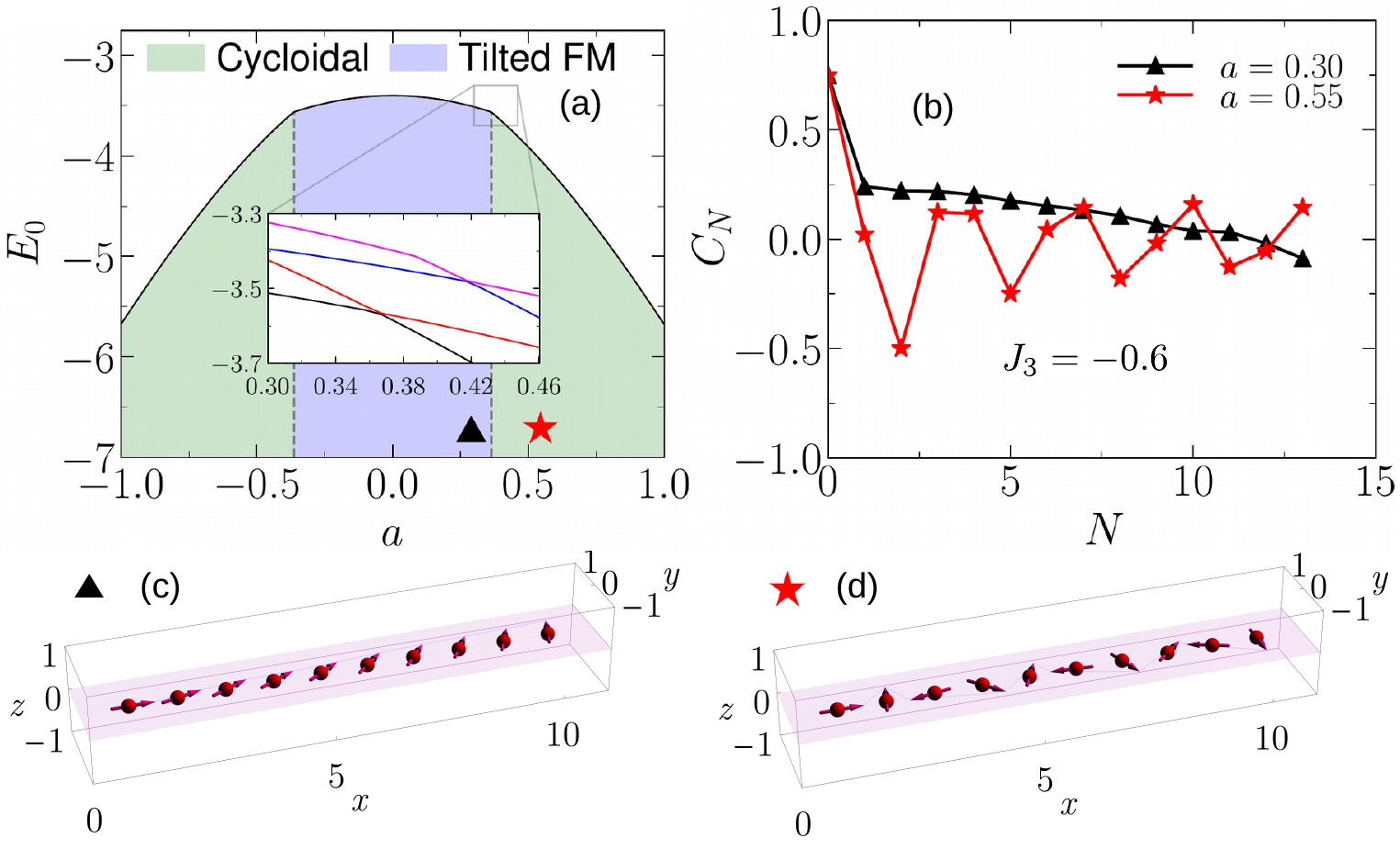}}   
	\caption{(a) Ground state energy vs $a$, the ratio between DM and  RKKY interactions. Blue and green regions represent the tilted FM and cycloidal phases, respectively, emerging from the FMLRO phase with $J_3=-0.6$ in Fig.~\ref{fig1}(a). Inset in (a) shows zoom-in of  region close to the transition  and includes the four lowest levels. Notice that the ground and first excited states cross at $a=a_c \approx 0.364$. (b) Longitudinal correlations vs $N$ for different $a$ values corresponding to black triangle and red star in (a). (c) and (d) view of spins in the tilted FM and cycloidal phases in panel (b), respectively. Cases shown for $L=14$.} 
	\label{fig1_2}
\end{figure}

Let us now turn on the DM interaction by making $a\neq 0$. As mentioned, there is a gradual tilting of spins away from either the DQLRO or FMLRO phases at $a=0$. Figure \ref{fig1_2}(a) shows $E_0$ vs $a$ for $J_3=-0.6$. We see  slope discontinuities in $E_0$ at $a = \pm a_c$,  indicating level crossings  separating distinct phases. One such level crossing is seen in the inset,  and includes the lowest few energy levels of the system \footnote{As only the $z$-component of the DM interaction is included in the Hamiltonian,  we restrict the calculation to the zero magnetization sector. Notice that $E(-a)=E(a)$ in Fig.~\ref{fig1}(e), reflecting the inversion symmetry in the Hamiltonian $x \leftrightarrow y$ while $a\rightarrow -a$.}. Figure \ref{fig1_2}(b) shows $C_N$ for $a=0.30$ (black) and for $a=0.55$ (red) on different phases. For $a$ in the blue area in \ref{fig1_2}(a), $C_N$ starts positive and slowly decreases for large separation  $N$. This signals the competition between DM  and RKKY interactions, as the  spins  are progressively tilted along the chain to minimize the overall energy, while keeping local near FM alignment. This situation is visualized in Fig.\ \ref{fig1_2}(c). Notice that the spins are tilted on the $xy$ plane along the chain, and tilt gradually more as the DM interaction $|a|$ increases. However, for $|a|>a_c$, the system collapses  suddenly into a very distinct phase, with $C_N$ showing an oscillatory behavior deceptively similar to the VBS+QLRO phase previously described \cite{Sandvik2010}, although with drastically different spin structure. As we will see below, this phase can be described as having a cycloidal quasi-long range order with sizable VSC\@.  The spin configuration for $a=0.55$  in Fig.\ \ref{fig1_2}(b) is shown in \ref{fig1_2}(d), which clearly exhibits cycloidal correlations on the $xy$-plane similar to those found in different systems \cite{Fishman2019}.  
Despite the small size ($L=14$) of the chains shown so far, they exhibit the main features of larger systems in this regime. We have investigated chains as large as $L=22$ with full diagonalization and observe the same general behavior. The transition points, although weakly, depend on the chain length $L$; a linear scaling fit to $L^{-1}$ yields $a_c \approx 0.417$ for $L\rightarrow \infty$.

\begin{figure}[t!]
	\centering
	\subfigure{\includegraphics[clip,width=3.4in]{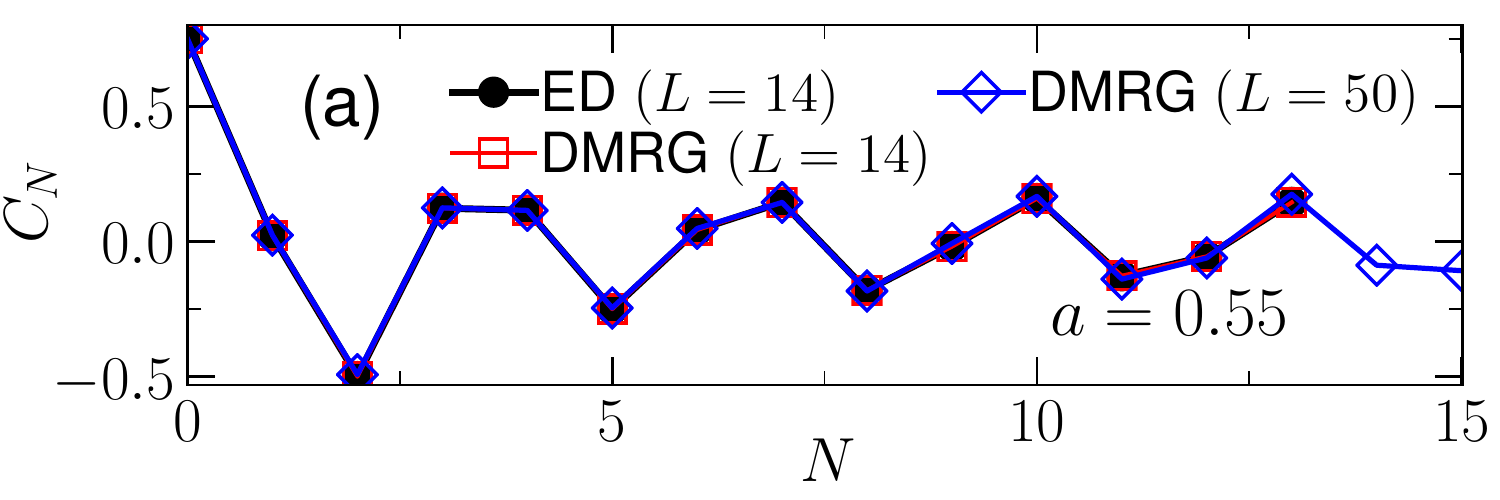}}    
	\vskip-0.3cm
	\subfigure{\includegraphics[clip,width=3.4in]{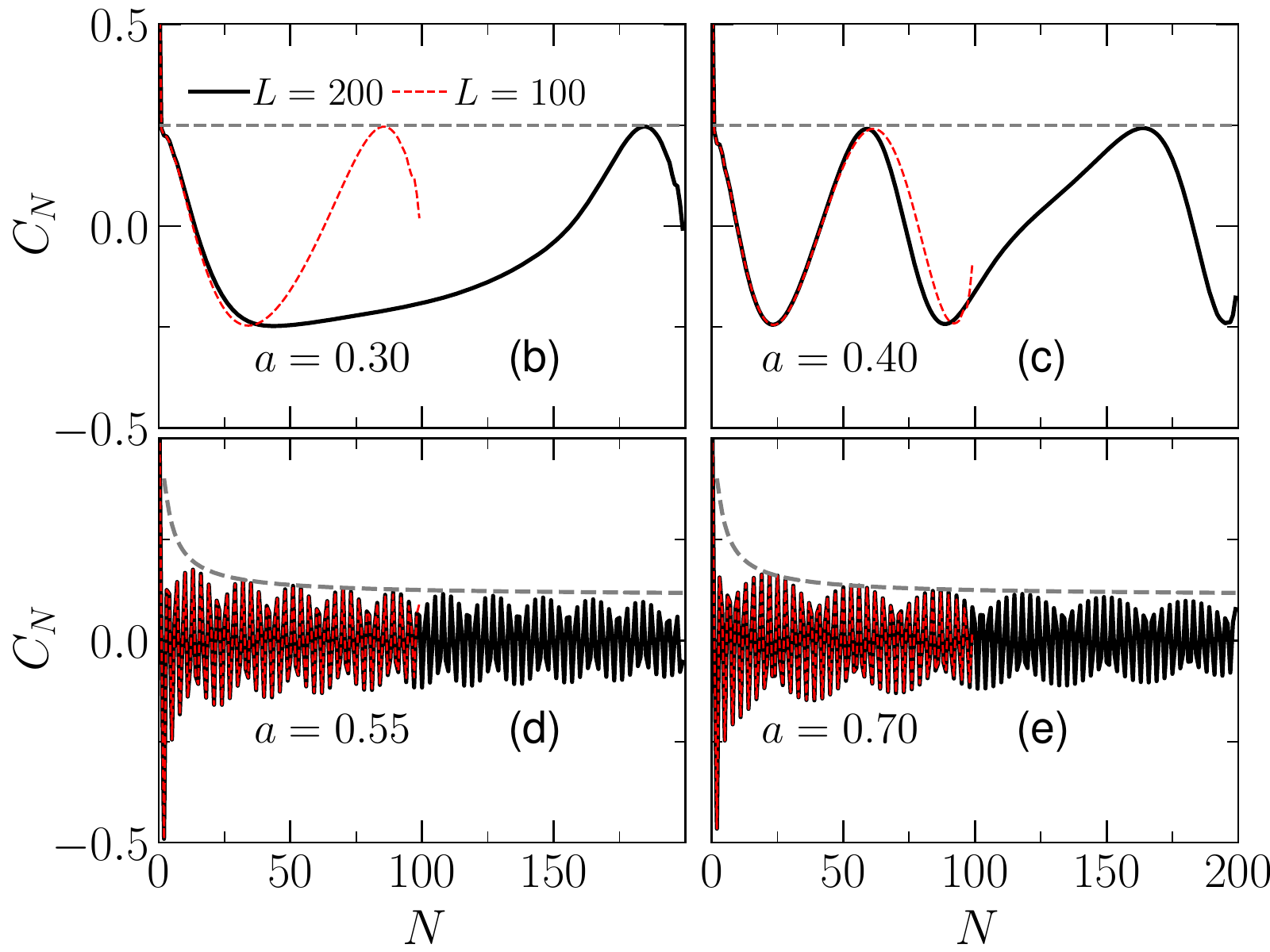}} 
	
	\caption{(a) Comparison of spin-spin correlations vs $N$ for ED and DMRG and different length $L$ for $a=0.55 (>a_c)$. There is remarkable agreement for $L=14$, and only a slight deviation for $L=50$, indicating weak finite size effects.  (b) and (c) show $C_N$ vs $N$ obtained with DMRG for $a=0.3$ and $0.4$, both $<a_c$, in the tilted FM phase. Red (dashed) and black lines correspond to $L=100$ and $L=200$. (d) and (e) show results for  $a=0.55$ and $0.70$, within cycloidal phase. Long-dash (gray) line in (b) and (c) corresponds to $f(N)=1/4$, while in (d) and (e) it is given by $f(N)=0.7\sqrt{\ln(N)}/N+0.11$. In all panels $J_3=-0.6$. 
		\label{fig5}} 
\end{figure}

 In Fig.\ \ref{fig5}(a) we compare the  spin correlation $C_N$ for $a=0.55$ and $J_3=-0.6$, corresponding to the cycloidal phase [star in Fig.\ \ref{fig1_2}(a)],  for both ED and DMRG calculations. Note again the perfect agreement for $L=14$, and only a small deviation in the last few points when compared with the DMRG results  for $L=50$. This suggests that finite size effects in this phase are not as important as above. Figure \ref{fig5}(b) and (c) show  $C_N$ for $a=0.30$ and $a=0.40$, both in the  tilted FM phase. Red (dashed) and black (solid) curves correspond to $L=100$ and $200$, respectively, demonstrating strong finite size effects for these $a < a_c$ values. Comparing both panels for a given $L$ we see that the results depend also strongly on $a$. Remarkably, there is no apparent decay in $C_N$, suggesting LRO where the local alignment of spins is nearly parallel to each other but slowly tilting away with a long wavelength which gets smaller as $a$ increases, as one would expect.

Figures \ref{fig5}(d) and (e) show results for $J_3=-0.6$ and  $a=0.55$ and $a=0.70$, both in the cycloidal phase. In this regime, the  $C_N$ results are nearly independent of $L$ (with some edge effects) and exhibit different  oscillating patterns depending on the value of $a$. In this regime, $C_N$ exhibits a slow decaying envelope along the chain, indicating QLRO. 
\begin{figure}[t!]
	\centering
	\subfigure{\includegraphics[clip,width=3.47in]{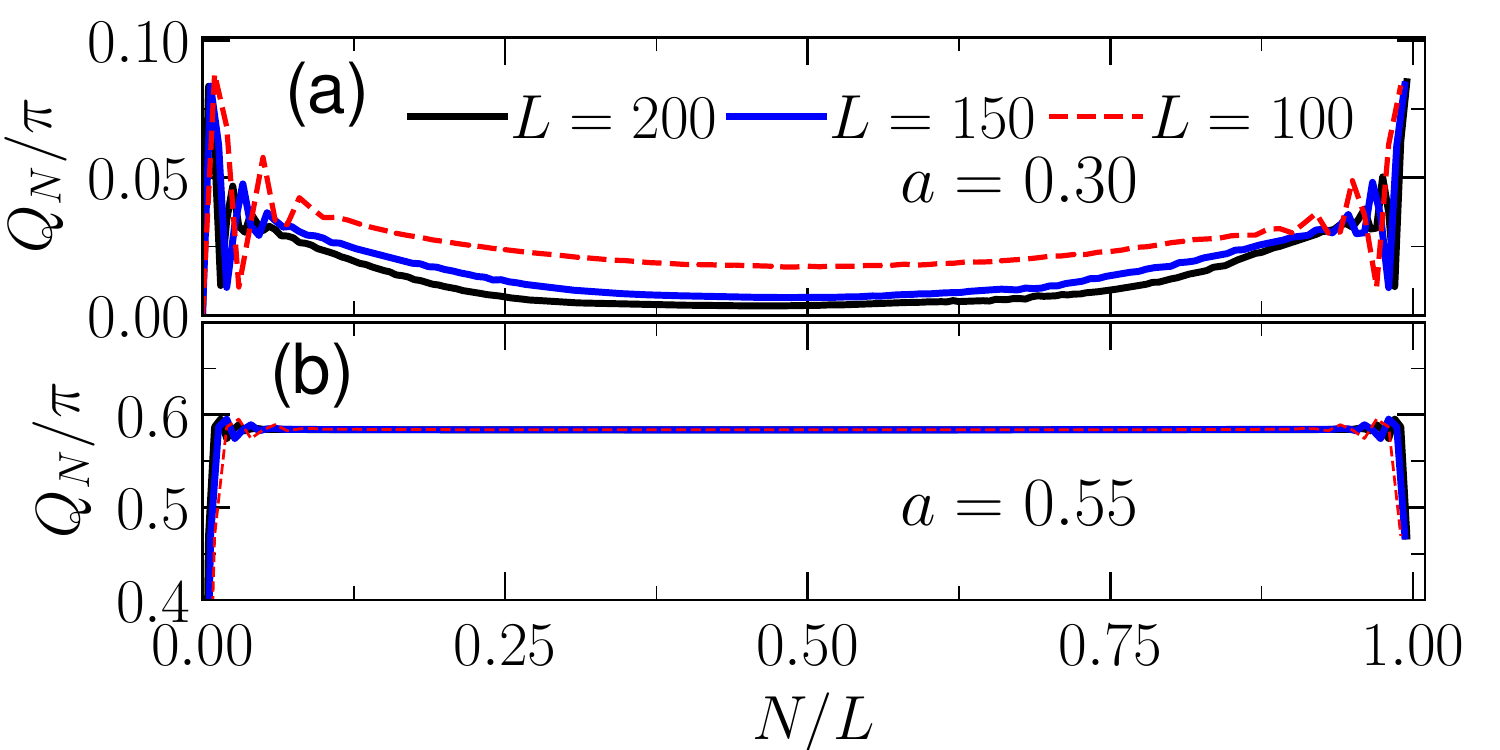}}
	\vskip-0.2cm	
	\subfigure{\includegraphics[clip,width=3.25in]{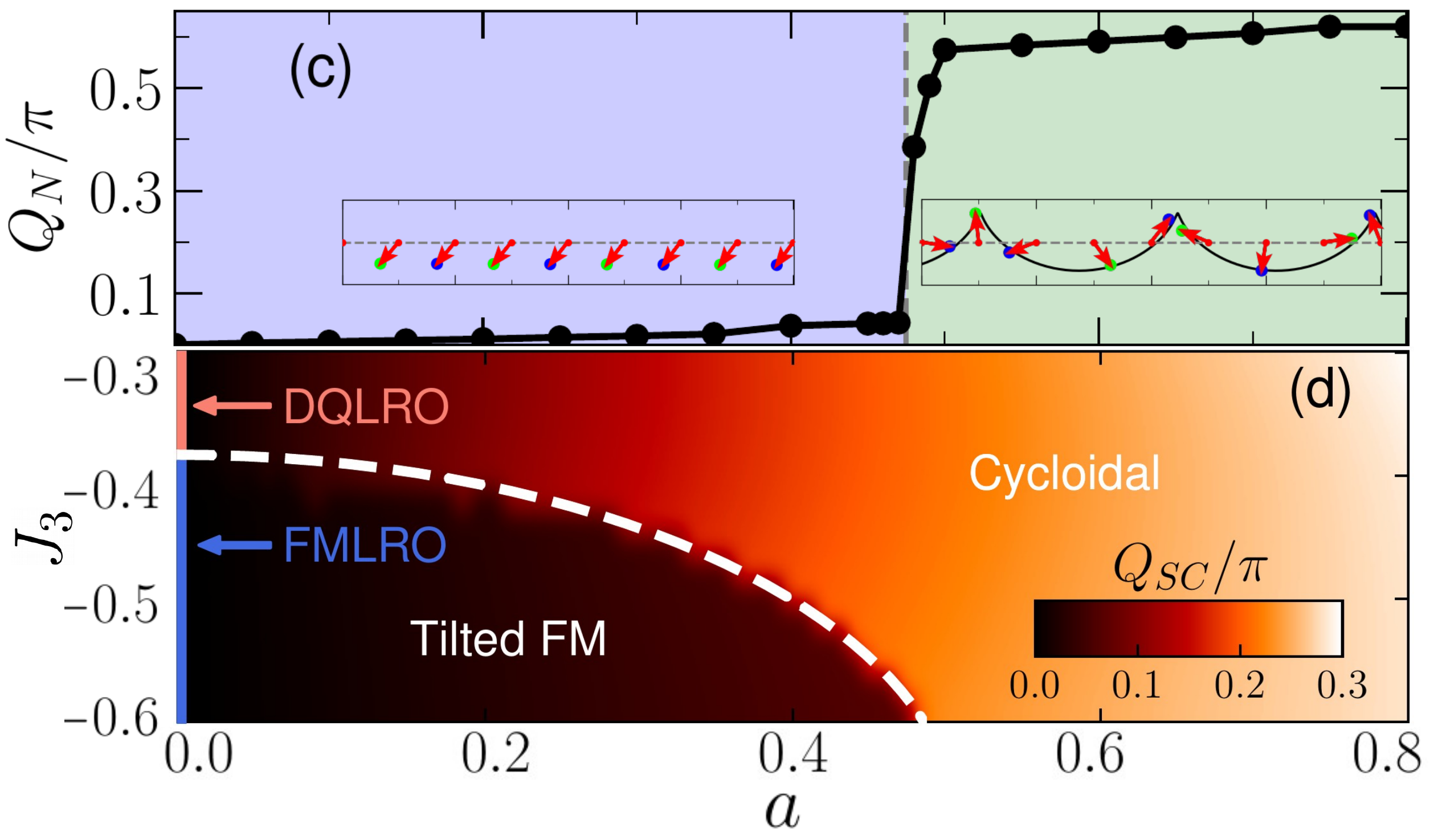}} 
	\caption{Spin rotation wavenumber $Q_N$ vs $N$ for (a) $a=0.30$  and (b) $a=0.55$, from Eq.~\eqref{angles} for different chain lengths $L=100$ (red), 150 (blue) and 200 (black).  Note that cycloidal phase in (b) is characterized by constant $Q_N$ (with variations on the edges) for all chain lengths. (c) $Q_N$ jumps across the transition between tilted FM (blue) and cycloidal (green) phases at $a=a_c$.  Results for $L=150$.  Insets show characteristic spin structures. $J_3=-0.6$ in panels (a) to (c). (d) Color map of the VSC wavenumber $Q_{SC}$.  Notice its gradual evolution as $a$ increases away from the DQLRO phase (for $a=0$ and $J_3>J_{3c}$).  In contrast, $Q_{SC}$ jumps from the tilted FM to the cycloidal phase at $a_c (J_3)$, indicated by white dashed line.} 
	\label{fig6}
\end{figure} 

We characterize the  spin structures in different regimes  by considering the direction of each spin along the chain as
$
\hat{\bf s}_N=\cos(\theta_N)\hat{\bf x}+\sin(\theta_N)\hat{\bf y}$,
%
where $\theta_N$ defines the orientation of the $N$-th spin  with respect to the $x$-axis. As the DM interaction in Eq.\ \eqref{H0} accounts only for the $z$-component of the cross product, the ground state requires all spins to be on the $xy$-plane. The coordinates of the tip of the vector $\hat{\bf s}_t$ with respect to the the first spin are given by the cycloid parametric equations, 
$
x_t=t+\cos(Qt) \quad  \mbox{and} \quad y_t=\sin(Qt)$.
%
The mean value of $Q$ can be determined from the accumulated relative phase of nearest neighbor spins
\begin{eqnarray}\label{angles}
Q_N=\frac{1}{NS^2}\sum_{i=1}^{N}\arccos\left(\langle {\bf S}_{i-1}\cdot {\bf S}_i \rangle \right),
\end{eqnarray}
where $\langle {\bf S}_{i-1}\cdot {\bf S}_i \rangle$ is the ground state  correlation function between adjacent sites.

Figure \ref{fig6} shows $Q_N$ vs $N$ for $a=0.30$ and $a=0.55$, on opposite sides of the phase boundary for $J_3=-0.6$; red, blue and black curves correspond to different size $L$.  Note that while $Q_N$ is constant for $a=0.55$ in the cycloidal phase (except for edge variations), it is substantially smaller (longer wavelength) and clearly affected by finite size for $a=0.30$, in the tilted FM phase \footnote{Notice that the cycloidal wavelength $\lambda=2\pi/Q_N$ ($\approx 3.44$) in Fig.\ \ref{fig6}(b) is in general not commensurate with the lattice, which results in the moire aliasing that produces the weakly oscillatory envelope in Fig.\ \ref{fig5}(d)-(e).}. As seen before, strong variation of $Q_N$ with $L$ is a reflection of the LRO in the tilted FM phase.  
It is important to note that the value of $Q_N$ jumps across the transition point $a=a_c(J_3)$, signaling the transition between the tilted FM and cycloidal phases.  Figure \ref{fig6}(c) shows such behavior (for $L=150$ and $J_3=-0.6$), as well as characteristic spin structures on both sides of the transition.

As mentioned before, the appearance of DM in the system favors the onset of 
a VSC in the ground state of the system.  We assess its spatial dependence from the corresponding accumulated value of the chirality
\begin{equation}
Q_{SC} = \frac{1}{NS^2}\sum_{j=1}^{N}\arcsin \left(\langle ({\bf S}_{j-1} \times {\bf S}_j ) \cdot \hat{z} \rangle \right).
\end{equation}
Interestingly, $Q_{SC}$ is found to follow closely $Q_N$ (ignoring edge effects) within a factor of two, and similarly seen to jump across the phase transition described, as long as $J_3$ places the system in the FMLRO phase at $a=0$.  However, for less negative $J_3$, where the system is in the dimerized phase (DQLRO), increasing $a$ produces only a gradual increase of $Q_{SC}$.  This interesting behavior is shown in Fig.\ \ref{fig6}(d), where the different phases identified above are labeled.   
Notice the sharp phase boundary indicated by the dashed white line.


The possibility of controlling the relative strength of different RKKY and DM interaction terms in a variety of systems and materials suggests that one can  exploit the ability to create a robust VSC of the system, and use that to drive information across the chain.  That the spin chirality can be turned on and off at will in a system would even suggest its use in reconfigurable spin transistor architectures.  
Our results complement the rich dynamics explored in experiments on spin chains with tunable magnetic interactions, and invite further investigations of the excitations using field theoretical techniques \cite{Book_Giamarchi,Hikihara2008,Garate2010}.  As experiments with ytterbium ion chains have demonstrated time crystals \cite{zhang2017observation}, iron atomic chains on different metal surfaces have been utilized for quantum information transfer \cite{PhysRevLett.108.197204}, or for the study of Majorana fermions \cite{Science2014NadjPerge,Science2017Jeon}, one could exploit the transition to a cycloidal phase as the one we discuss here.  Long range DM interactions can clearly produce a rich phase diagram even in 1D spin systems, which we expect would be exploited in different experiments.

We are thankful with D. Mastrogiuseppe and B. Bravo for very fruitful interactions and discussions at the start of this project.
We acknowledge support from NSF grant DMR 1508325; E.V. acknowledges the support of the Glidden Visiting Professor program at Ohio University.

%

\end{document}